\documentclass[showpacs,amsmath,amssymb,twocolumn,floatfix,prl]{revtex4}

\usepackage[dvips]{graphicx}
\usepackage{epsfig}
\input{epsf}

\begin{document}

\title{
Directed transport born from chaos in asymmetric antidot structures
}

\author{A.D.~Chepelianskii}
\affiliation{Ecole Normale Sup\'erieure, 45, rue d'Ulm, 
75231 Paris Cedex 05, France
}

\date{September 27, 2005}

\begin{abstract}
It is shown that a polarized microwave radiation creates directed transport 
in an asymmetric antidot superlattice in a two dimensional electron gas. 
A numerical method is developed that allows to establish the 
dependence of this ratchet effect on several parameters relevant 
for real experimental studies. 
It is applied to the concrete case of a semidisk Galton board 
where the electron dynamics is chaotic in the absence of microwave driving.
The obtained results show that high currents 
can be reached at a relatively low microwave power.
This effect opens new possibilities for microwave control of transport 
in asymmetric superlattices.  
\end{abstract}
\maketitle

\section{Introduction}

The appearance of a directed transport induced by radiation 
in asymmetric systems is known as the photogalvanic effect. 
By this effect the radiation creates charge transport 
in the bulk of the asymmetric structure in absence of any 
applied {\it dc}-voltage. 
The theoretical investigations of this phenomena have been 
started almost 30 years ago in Refs. \cite{entin,belinicher}.  
The interest to this subject has been renewed recently 
with the studies of ratchets that appear when a system 
is displaced from thermal equilibrium by a periodic variation of system parameters 
(for reviews see Refs. \cite{hanggi,reimann}).
One of surprising properties of ratchets is that directed transport 
can emerge in presence of a zero mean force. 
This phenomenon has a generic origin and appears in various physical systems 
including  vortices in Josephson junction arrays 
\cite{mooij,nori,ustinov}, cold atoms \cite{grynberg},
macroporous silicon membranes \cite{muller}, 
microfluidic channels \cite{ajdari} and other systems. 

Nowadays technology allows to prepare artificial antidot superlattices 
in semiconductor heterostructure with two dimensional electron gas (2DES).
The conduction properties of these samples has been tested in experiments 
\cite{weiss,kvon} which showed an important contribution of periodic orbits. 
The structure of these superlatticies is similar to a periodic lattice
of rigid disks placed on a plane. 
Such structures are known as Galton boards \cite{galton} 
or Lorentz gas. According to the mathematical results of Sinai the dynamics 
on such a lattice is completely chaotic \cite{sinai}.
The theoretical studies \cite{geisel}  
performed to understand these experiments showed 
that the chaotic classical dynamics and periodic orbits significantly affect 
the conduction properties in such superlatticies.

The effects of microwave radiation on the conduction properties of antidot 
superlattices has been addressed in experiments \cite{kvon1}.
However in these structures due to the symmetry of the superlattice 
the ratchet effect was forbidden. Asymmetric mesoscopic structures 
under external periodic driving have been addressed 
in the experiments \cite{linke}. 
However zero mean force ratchet was absent due to low frequency of driving
which was essentially adiabatic \cite{linke1}.

The recent theoretical studies of dissipative 
transport in asymmetric superlatticies 
showed that microwave radiation induces directed transport in such systems 
(zero mean force ratchet) \cite{second,cristadoro}.
These works were mainly performed for a semidisk Galton board which 
is obtained from the usual Galton board of rigid disks by replacing 
each disk with semidisk oriented in one fixed direction (see Fig. 1).
In Ref.~\cite{second} the model of 
a friction force $\mathbf{f}_f = - m \gamma \mathbf{v}$ 
with constant friction coefficient $\gamma$ has been used for a particle 
of mass $m$ moving with velocity $\mathbf{v}$. 
In Ref.~\cite{cristadoro} the case of particles 
in a Maxwell thermostat at temperature $T$ was considered. It was shown 
that the thermostat creates an effective friction coefficient $\gamma$ 
which depends on the microwave field strength and 
the temperature of the thermostat. 
However the most interesting case is the Fermi-Dirac thermostat since it 
describes the experimental conditions 
of antidot superlatticies with 2DES \cite{rammer}.
Until now no numerical studies were performed in this regime. 
Only theoretical estimates have been proposed in Ref.~\cite{cristadoro}. 
Their validity was never checked and remains questionable.

\begin{figure}[t!]  
\vglue +0.5cm
\epsfig{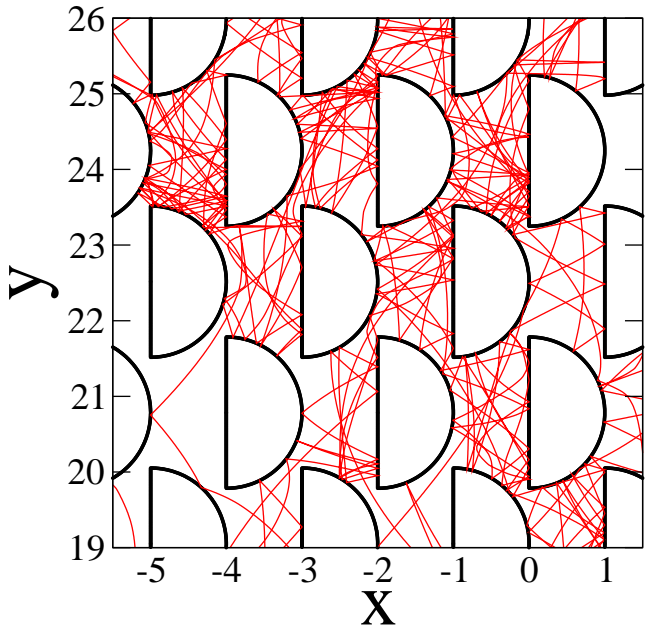}
\vglue -0.0cm 
\caption{(color online) Example of the semidisk Galton board with one 
chaotic trajectory in the 
region $-5.5 \le x \le 1.5$ and $19 \le y \le 26$.
Semidisk scatterers, separated by a distance $R = 2$, are shown in black.
The trajectory is displayed in red / gray.
The strength of microwave radiation and its frequency are 
$f = 5$, $\omega = 1$, polarization angle $\theta = 0$
and temperature $T = 0.1 E_F$. Here and in other figures $E_F = 10$
and disk radius $r_d=1$.
The Metropolis algorithm parameters 
are $\Delta E / E_F = 0.075, \Delta t = 0.005$. 
}
\label{fig1}       
\end{figure}

In this work I develop a numerical method which allows to study 
the directed transport induced by polarized microwave radiation 
$\mathbf{f} = f (\cos \theta, \sin \theta) \cos( \omega t ) $
in 2DES at various values of the Fermi energy $E_F$ and temperature $T$
(here $\omega$ is the radiation frequency and $\theta$ is the polarization 
angle with respect to the $x$ axis in Fig. 1).
On the basis of this method I performed 
extensive numerical studies which allowed to establish
the dependence of ratchet flow velocity $v_f$ 
on various system parameters 
including $T$ and $E_F$. 
Contrary to the estimates proposed in Ref.~\cite{cristadoro}
the dependence on $T$ is weak when $T \ll E_F$. 
The obtained results allow to predict typical values of currents in 
asymmetric antidot superlattices. 

The paper is organized as follows, in Section II the description of the 
model and of the numerical method is presented, the results are described 
in Section III, and conclusion is given in the last Section. 

\section{Model description}

The geometry of the model is represented by a 
Galton board of semidisks which form a 
two-dimensional hexagonal lattice
as represented in Fig. 1. 
The radius of the semidisk is $r_d$ and the distance 
between the disk centers is $R$.
A particle with mass $m$ moves under the action of 
a force $\mathbf{f}$ created by a polarized microwave radiation.
The collisions with the semidisks are elastic. 
In the numerical simulations I use dimensionless units 
with $m = r_d = 1$ and the Fermi energy $E_F = 10$ (thus the Fermi velocity $V_F = \sqrt{20}$).
In order to convert the numerical results obtained with this 
units, one should determine the dimensionless ratios: 
for example in Fig.~1 the frequency $\omega = 1$ and field strength $f = 5$ 
corresponds to dimensionless values $\tilde{\omega} = \omega r_d / V_F = 1 / \sqrt{20}$.
and $\tilde{f} = f r_d / E_F = 1 / 4$.

It is assumed that particles are non interacting 
but that the contact with a thermostat creates 
the Fermi-Dirac distribution $f_F(E) = 1 / (E_F [ \exp((E - E_F)/T) + 1 ] )$ 
at temperature $T$ and Fermi energy $E_F$, 
where E is a particle kinetic energy.
Here the energy distribution for one particle $f_F(E)$ is normalized 
by the condition $\int f_F(E) dE = 1$.
This one particle distribution gives also the result for 
many particles by simple rescaling by the number of particles
as it is usually done for 2DES (see \cite{rammer} p.193). 
Such a situation corresponds 
to experiments with 2DES in antidot lattices 
similar to those of Refs. \cite{weiss,kvon}.
To keep particles in a thermal equilibrium 
with the Maxwell distribution it is possible to use 
various methods including the Nos\'e-Hoover thermostat
used in Ref.~\cite{cristadoro}. However for the Fermi-Dirac 
thermal distribution this method is not appropriate 
and a new approach should be developed. 
Indeed the Nos\'e-Hoover equations are constructed in such a way 
that they give the Maxwell distribution 
in particle velocities \cite{hoover,cristadoro}.
They should be significantly modified to generate the Fermi-Dirac distribution 
and until now this problem has not been addressed yet. 
The first attempt by the authors of Ref.~\cite{cristadoro} did not 
succeed in achieving a good convergence to the Fermi-Dirac distribution \cite{privecristadoro}.

\begin{figure}[t!]  
\vglue 0.5cm
\epsfig{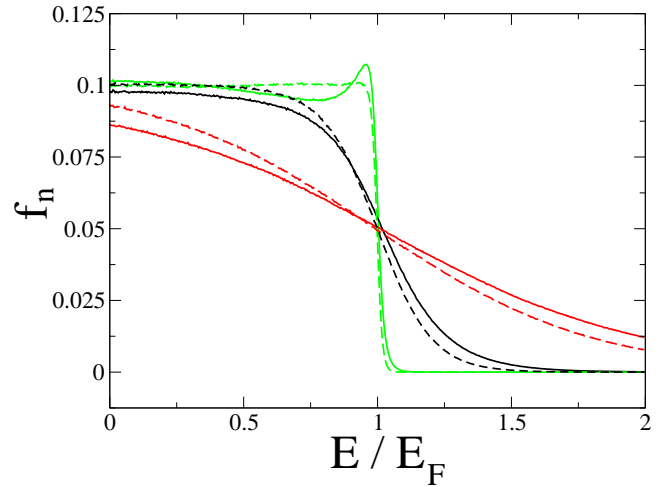}
\caption{(color online) 
Energy distribution functions 
$f_n(E)$ obtained numerically for different 
field amplitude values $f$ and different temperatures $T$.
For the green, black and red 
curves the temperatures are $T / E_F = 0.01$, $0.1$, and $0.4$ respectively
(at $E / E_F = 1.1$ the order of curves is from bottom to up).
The applied force $f$ is zero for the dashed curves
and $f = 5$ for the full curves ($\omega = 1$, $\theta = 0$, $R = 2$).
At $f = 0$ the dashed curves coincide with the Fermi-Dirac distribution
$f_F(E)$ at corresponding temperature $T$.
The Metropolis algorithm parameters are the same as in Fig. 1.
}
\label{fig2}       
\vglue +1.0cm
\end{figure}

My approach is inspired by the successful Monte Carlo 
method adapted to simulate  numerically the transport 
properties in semiconductor devices \cite{jacoboni}.
To obtain a stable Fermi-Dirac thermalization
of particles on the semidisk Galton board (see Fig. \ref{fig1}) 
the following procedure has been applied: 
(i) the equations of motion were integrated exactly 
on the time interval $\Delta t$ using the analytical 
solution; 
(ii) after that the particle energy is 
changed from its value $E$ to another value 
in the interval $(E - \Delta E, E + \Delta E)$
without changing the direction of the particle momentum.
The choice of this value 
is governed by the Metropolis algorithm \cite{metropolis}
which imposes the convergence to the Fermi-Dirac distribution $f_F$.
Namely, a random value $E'$ is chosen in  $(E - \Delta E, E + \Delta E)$,
with probability $\min( f_F(E') / f_F(E), 1)$ 
the algorithm sets $E = E'$, otherwise $E$ remains unchanged. 
Afterward the algorithm returns to step (i).
The times when collisions with semidisks occur 
are found with the precise Newton algorithm as in Ref.~\cite{second}.
The step $\Delta E$ can be considered as a thermalization step 
which determines the rate of convergence to the equilibrium distribution 
$f_F$. 

The Metropolis algorithm described above gives convergence 
to the equilibrium distribution $f_F$ in the absence of 
microwave radiation. The examples of steady state distributions 
at different temperatures $T$ are shown in Fig. 2.
The proximity of the numerically obtained distribution $f_n(E)$ 
to the theoretical steady state $f_F(E)$ can be characterized
by the dimensionless mean square deviation 
$\sigma =  E_F \int (f_n(E) - f_F(E))^2 dE$.
This quantity remained small in all numerical simulations at $f = 0$ 
showing a good convergence to the Fermi-Dirac distribution (e.g 
for the cases of Fig.~2: $\sigma = 1.8 \times 10^{-5}, 2.3 
\times 10^{-5}$ and $8.3 \times 10^{-5}$ 
for $T / E_F = 0.4, 0.1, 0.01$ respectively).

\begin{figure}[t!]
\vglue +0.5cm
\epsfig{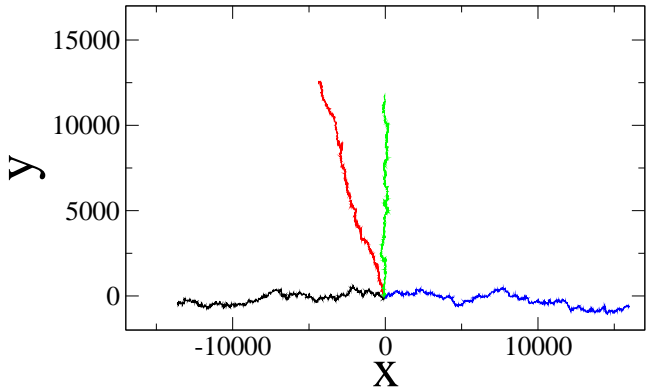}
\vglue -0.0cm
\caption{(color online) Directed transport 
for one trajectory at various polarizations of radiation 
$\theta = 0$, $\pi / 5$, $\pi/4$, $\pi/2$ (from left to right clockwise), 
other parameters are set as in Fig. 1. 
}
\label{fig3}
\end{figure}

However, the introduction of the microwave radiation modifies 
the distribution $f_n$ which depends on $f$ and other system parameters. 
This is clearly seen in the typical cases presented in Fig. 2. 
For relatively high temperatures the distribution $f_n$ remains 
a smooth monotonic function of energy whereas at low $T$ the 
microwave field creates a characteristic peak near the Fermi energy $E_F$.  
As a result the developed numerical method allows to study 
the transport created by microwave radiation in an asymmetric 
semidisk Galton board in the stationary regime. To be close 
to realistic experimental situations additional random scattering 
has been introduced to take into account
the effect of impurities. Namely after time $\tau_i$ the direction 
of particle momentum is changed randomly (angle changes in the
interval  $[0, 2 \pi]$).
In the majority of cases studied the value $\tau_i$ was kept 
sufficiently large ($\tau_i V_F / r_d > 1000$) and did not influence 
the stationary transport properties (the dependence on $\tau_i$
will be discussed in the next Section). Here $V_F = \sqrt{2 E_F}$ is the Fermi 
velocity.

This stationary regime,
which sets in presence of radiation,
clearly shows the photogalvanic (ratchet) effect 
with directed transport born from chaos as it is shown in Fig.~3 
for different polarization angles $\theta$. 
I would like to stress that this effect originates from the bulk 
of the sample; it is known that radiation can lead to the appearance 
of electron-hole pairs, which may diffuse to different contacts 
because of the electric field in the depletion region of the 
junction, however in this case the effect is due to the 
contacts and not to the bulk of the sample. 
For $\theta = 0$
the transport is directed to the left (negative direction on the $x$ axis),
while for $\theta = \pi / 2$ the transport is oriented to the right.
The polarization dependence is similar to that obtained in previous 
works \cite{second,cristadoro} and is well described by 
the relation $\psi \approx \pi - 2 \theta$ where $\psi$ is the angle 
between the transport direction and the $x$ axis. 
In this way the average velocity of transport can be written 
as $\mathbf{v_f} = v_f (\cos \psi, \sin \psi)$.
It is calculated by following a single trajectory for a long 
typical time $t V_F / r_d \sim 10^7$. It was also checked that the 
averaging over an ensemble of few tens of different trajectories 
gives statistically the same result. 

\begin{figure}[t!]
\vglue +0.5cm
\epsfig{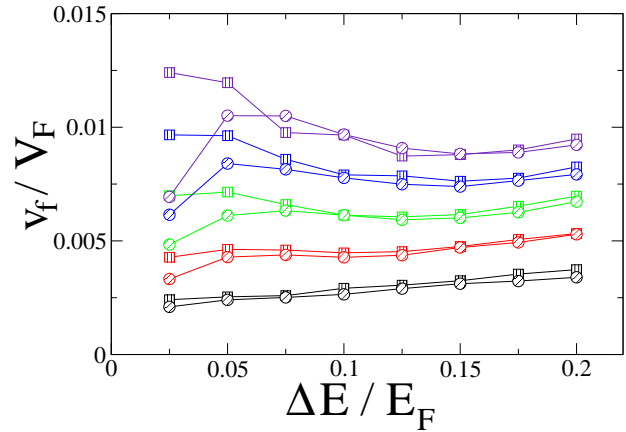}
\vglue -0.0cm
\caption{(color online) The ratio of flow velocity 
$v_f$ to the Fermi velocity $V_F$ 
as a function of the thermalization step $\Delta E$ 
for a fixed $\Delta t = 0.005$. 
The different curves correspond to values of $f$: 
$f = 7.0$, $6.0$, $5.0$, $4.0$, $3.0$ with respective colors: 
violet, blue, green, red, black
(from top to bottom at $\Delta E / E_F$ = 0.2).
The symbols mark the temperature $T / E_F = 0.01$ (squares)
and $T / E_F = 0.1$ (circles).
Other parameters are as in Fig. 1.
}
\label{fig4}       
\end{figure}

Special checks have been made to ensure that the ratchet velocity $v_f$ 
does not depend on the Metropolis algorithm parameters.
An example of such checks is shown in Fig. 4 where the thermalization 
step $\Delta E$ was changed by  one order of magnitude. 
In spite of this variation the value of $v_f$ remains
stable. It was also checked that 
the resulting ratchet velocity is not sensitive 
to variation of $\Delta t$, e.g a variation
of $\Delta t$ by more than one order of magnitude 
gave no variation of $v_f$ within a 5\% accuracy.
Physically it is possible to say that the relaxation time 
to the equilibrium $\tau_{rel}$ is approximately given by the relation 
$\tau_{rel} \sim \Delta t E_F^2 / \Delta E^2 $. 
Thus, the numerical checks above can be physically interpreted 
as the independence of the ratchet velocity on the variation 
of the energy relaxation timescale $\tau_{rel}$
which varied by two orders of magnitude. A similar effect has been seen 
with the Nos\`e-Hoover thermostat for the Maxwell equilibrium 
distribution \cite{cristadoro}. 
This result is also in a qualitative argument with the 
theoretical arguments given in Ref.~\cite{entin,belinicher}
according to which $\tau_{rel}$ does not directly affect $v_f$.
In my further simulations the Metropolis algorithm parameters are set 
to typical values $\Delta E / E_F = 0.075, \Delta t = 0.005$.
The data shown in Fig. 4 clearly demonstrate that $v_f$ grows 
with the radiation strength $f$. At the same time $v_f$ is 
not very sensitive to the variation of temperature $T$.  
Detailed studies of parameter dependence of $v_f$ are presented
in the next Section.

The results presented above show that the developed 
algorithm allows to simulate non interacting electrons 
at thermal equilibrium with the Fermi-Dirac distribution.
The microwave driving is assumed to be relatively weak 
so that it gives only small deviations from the unperturbed distribution,
this fact is clearly illustrated in Fig.~2. 
In this regime the perturbed distribution of particles 
is determined by the unperturbed distribution $f_F(E)$ and 
the microwave field, the effect of the latter is exactly taken into account 
by the Hamiltonian equations of motion. 
Therefore the developed algorithm should correctly describe 
the non equilibrium steady state distribution that emerges 
under mircrowave driving. 
This is indirectly confirmed by the fact that the directed 
transport is not sensitive to the thermalization step of the 
Metropolis algorithm (see Fig.~4).
In a sense the Metropolis steps combined with the Hamiltonian 
equations of motion give the solution of the kinetic Boltzmann
equation in the presence of microwave driving. 

The above arguments should be also valid for another 
unperturbed thermal distribution, e.g the Maxwell distribution 
$f_M(E) = \exp(-E / T) / T$. This case was analyzed in Ref.~\cite{cristadoro}
on the basis of Nos\'e-Hoover equations. 
In fact the
Metropolis algorithm had been invented to treat the Maxwell
thermal equilibrium \cite{metropolis}.
Thus, I made numerical tests 
with the Metropolis algorithm for the Maxwell distribution. 
The obtained results reproduce the functional dependences found in 
Ref.~\cite{cristadoro} (see Eqs.~(4,5) there) with approximately 
the same values of the numerical constants.
This gives independent confirmation that the Metropolis algorithm 
treats correctly a weak external perturbation that drives the 
system out of equilibrium. It also shows that various thermal 
distributions can be treated by this method.

The model I described assumes that the electrons are non interacting.
To be valid it requires that the Coulomb interaction between electrons in 2DES
$E_{ee} \approx e^2 \sqrt{\pi n_e}/\epsilon_r$ 
is small compared to the kinetic energy given by 
$E_F = \pi n_e \hbar^2 / m$. 
Here $n_e$ is the electron density, $\epsilon_r$ is the dielectric constant, 
$e$ is the electron charge and $m$ is the effective electron mass.
Thus the effective strength of interaction is characterized by 
the dimensionless parameter $r_s = E_{ee}/E_F$ \cite{sarachik}
which should be small. Its value for experimental 2DES obtained 
in $GaAs/AlGaAs$ heterostructure with electron densities 
$n_e \approx 10^{12} cm^{-2}$, 
an effective electron mass $m \approx 0.065 m_e$ 
and dielectric constant $\epsilon_r = 13$  
is approximately $r_s \sim 1$.
At such values the interaction between quasiparticles is considered 
to be weak and is usually neglected in a first approximation 
\cite{rammer,sarachik},
for example the Wigner crystal typically appears at $r_s \approx 37$.

At such densities $n_e$ inside a cell of size $S = 1 \mu m^2$ the 
quantum level number of an electron at the Fermi energy is 
$N_F = n_e S \sim 10^4$. Therefore electrons are in a deep semiclassical 
regime and the classical Monte Carlo approach used above is well 
justified, see also \cite{jacoboni}.

\section{Numerical results}

I have investigated the dependence of $v_f$ on several 
system parameters which are relevant for a realistic 
experiments with 2DES in antidot lattices. 
Among them are the temperature $T$, 
the microwave field strength $f$ and the microwave frequency 
$\omega$. The effects of geometry are studied by 
changing the lattice constant $R$ that allows to choose 
the optimal regime where the photogalvanic effect is 
stronger. The effects of impurity scattering is 
modeled by variation of scattering time $\tau_i$
that gives insight on the stability of the effect 
in respect to experimental imperfections. 
At last the effect of magnetic field is also analyzed.

\vglue 0.0cm
\begin{figure}
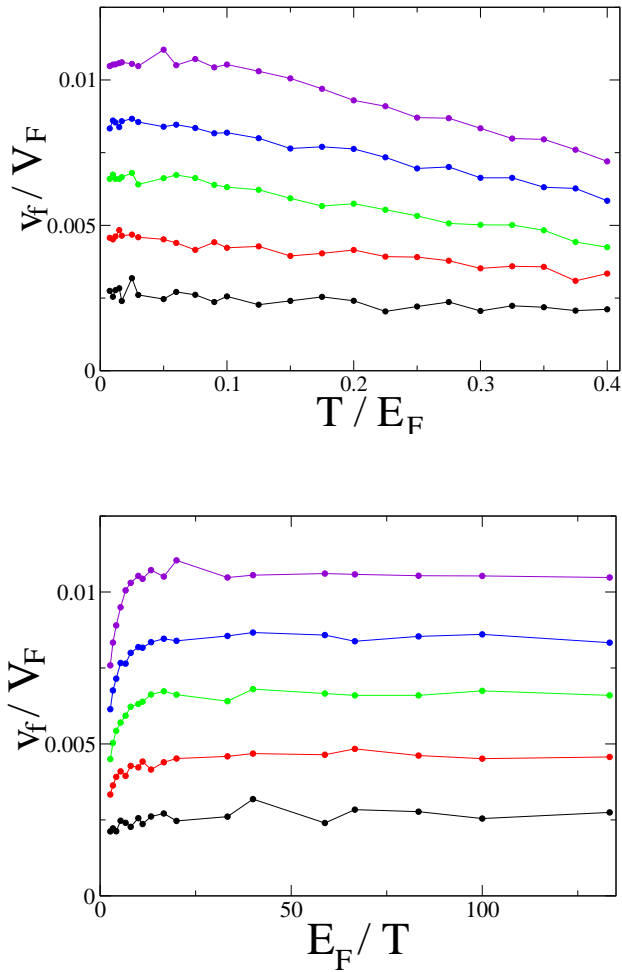

\vglue 0.5cm
\epsfig{file=fig5a.eps,width=3.2in}
\vglue +1cm
\epsfig{file=fig5b.eps,width=3.2in}
\label{fig5a}
\caption{(color online) Top panel: dependence of the rescaled flow velocity $v_f / V_F$ on the 
rescaled temperature $T / E_F$, 
different curves correspond to $f = 7.0$, $6.0$, $5.0$, $4.0$, $3.0$
(from top to bottom). Bottom panel: the same data are shown as a function of $E_F / T$. 
Here $\omega = 1$, $R = 2$, $\theta = 0$. 
}
\end{figure}

The temperature dependence for a typical set of parameters 
is given in Fig. 5. The obtained numerical data show that there is 
a weak drop of the ratchet velocity $v_f$ with the increase of 
temperature $T$. However in the regime with $T \ll E_F$, 
$v_f$ is practically temperature independent.
This dependence is preserved for various radiation 
strengths $f$. 
The velocity of transport increases with the growth of $f$.
A detailed study of the effect dependence on the microwave
field $f$ is presented in Fig. 6.
It shows that the dependence on $f$ is
quadratic in the region $T \ll E_F$, and temperature 
independent in a large interval of field strength. 
At higher $f$ a deviation 
from the quadratic dependence starts to be visible,
this deviation starts earlier at high temperatures. 
Thus on the basis of the results presented in Figs. 5,6  
it is possible to conclude that at $T \ll E_F$ 
and in the limit of weak driving the flow velocity is 
described by the relation :
\begin{equation}
v_f / V_F = C (r_d f / E_F)^2 
\label{eq1}
\end{equation}
Here the dimensionless factor $C$ may depend on the 
microwave frequency, lattice geometry, and impurity 
scattering time. However it is independent of $T$ and $f$. 
For $R = 2$, $\omega r_d / V_F \ll 1$ and $\tau_i V_F / r_d \gg 1$ 
the obtained data give $C = 0.129 \pm 0.002$.

\vglue 0.0cm
\begin{figure}
\vglue 0.5cm
\epsfig{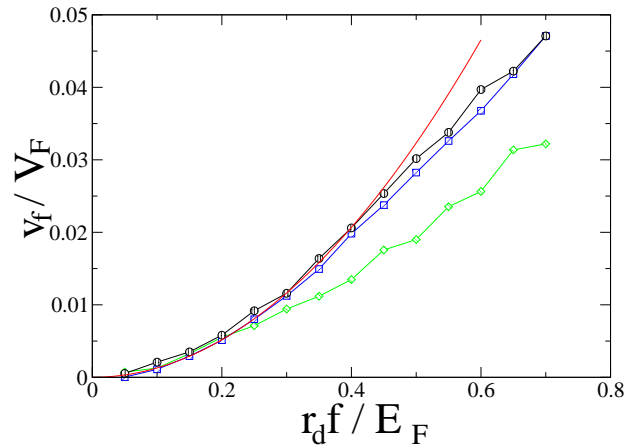}
\vglue -0.0cm
\caption{The rescaled flow velocity $v_f/V_F$ as a function of rescaled 
applied force $r_d f / E_F$  for different temperatures: 
$T / E_F = 0.4$ (green diamonds), $0.1$ (blue squares), 
$0.01$ (black circles). 
The red (full gray) curve shows a parabolic fit of data 
$v_f / V_F = C \; (r_d f / E_F)^2$ with $C = 0.129 \pm 0.002$ 
at $T / E_F = 0.01$
(the fit is done in the interval $[0,0.45]$).
Here $\omega = 1$, $R = 2$, $\theta = 0$.
}
\label{fig6}       
\vglue +0.5cm
\end{figure}

The dependence (1) is qualitatively different from
the theoretical estimates proposed in Ref.~\cite{cristadoro}.
To understand the origin of this difference I remind 
the main elements of estimates given in Refs.~\cite{second,cristadoro}.
They are based on the fact that the microwave 
radiation produces diffusive energy growth of electron 
energy in time with the rate : $D_E = (\delta E)^2 / \delta t$.
Here $\delta E$ is the energy variation after a time $\delta t$. 
In the limit of low frequency driving it is possible to write 
$D_E \sim (\dot E)^2 \tau_c \sim f^2 V_F l$ where $E$ is 
the particle energy and $l$ is the mean free path which is 
$l \sim R^2 / r_d$. If a particle would experience a friction 
force $\mathbf{f}_f = - m \gamma \mathbf{v}$, the 
diffusion in energy would give energy variation 
$(\delta E)^2 \sim D_E / \gamma$. In Ref.~\cite{cristadoro} it 
was assumed that $\delta E$ is fixed by the thermal distribution 
so that $\delta E \sim T$. Thus the statistical stationary 
distribution imposes an effective friction with coefficient 
$\gamma \sim D_E / T^2$. This relation is important 
because the ratchet velocity is given by relation 
$v_f \sim l \gamma$ in the limit of weak friction as it has 
been shown in Ref.~\cite{second} by extensive numerical simulations.
For the Maxwell distribution this gives $v_f \sim (l f)^2 / T^{3/2}m^{1/2}$ 
since in this case $D_E \sim f^2 (T/m)^{1/2} l$, where it is used 
that the thermal velocity $v \sim \sqrt{T/m}$.
The extension of the above arguments to the Fermi-Dirac distribution 
led the authors of \cite{cristadoro} to the conclusion that  
$v_f / V_F \sim C (f l / T)^2$. 
However this expression is drastically different from Eq.~(1) 
according to which there is no temperature dependence in the 
limit $T \ll E_F$. This contradiction can be resolved if 
the variation of particle energy induced by the radiation 
is of the order $\delta E \sim E_F$ (and not $\delta E \sim T$).
Indeed in the free electron model the particle can move 
in the whole energy interval defined by the Fermi energy 
whereas in the estimates in Ref.~\cite{cristadoro} it was 
assumed that the particle can move only in the narrow 
thermal layer near the Fermi surface. 
In the free particle model it is therefore rather natural 
that the particle energy variation is $\delta E \sim E_F$,
that leads to the result of Eq. (1).

The fact that the free electron model remains valid in the 
presence of microwave driving can be also understood 
from the following arguments. For non interacting electrons 
the Hamiltonian is given by the sum of one particle 
operators (microwave driving is also one particle operator). 
Hence, the many particle state
(wave function or density matrix)
is obtained simply from one particle states
by antisymmetrization. Thus,
 the Pauli principle can be taken into account 
by averaging the final one particle results over 
the Fermi-Dirac distribution. This statement also explains 
the validity of the classical kinetic Boltzmann equation 
for the description of transport properties of metals.  
It is demonstrated more rigorously for a non interacting 
Fermi gas in (Chap. 5 in Ref.~\cite{rammer} and Ref.~\cite{sturman}). 
As a consequence the Pauli blockade does not appear for non interacting 
particles and the arguments presented in Ref.~\cite{cristadoro}
are not valid at least for weak $r_s$ values.

The arguments given above allow to understand the physical 
origin of the relation between the ratchet velocity 
and effective friction coefficient induced by microwave 
radiation that perturbs the system out of thermal equilibrium.
Another approach has been developed recently in Ref.~\cite{entin2}
using perturbation theory for the Boltzmann kinetic equation 
in the limit of weak radiation and weak density of randomly 
distributed asymmetric scatterers. The model proposed
in Ref.~\cite{entin2} is rather different from the one considered here, 
e.g. scatterers are distributed randomly on the plane,
their density is required to be small and impurity 
scattering is necessary for the regularization of the model.
However in spite of these differences the global dependence 
of ratchet velocity on Fermi energy $E_F$ and radiation strength 
$f$ is the same as in Eq. (1).

\begin{figure}[t!]  
\vglue 0.5cm
\epsfig{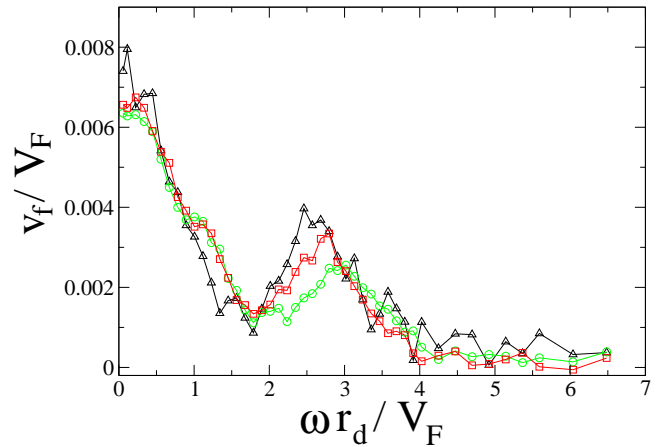}
\vglue -0.0cm 
\caption{(color online) Dependence of rescaled flow velocity 
$v_f / V_F$ on the rescaled microwave frequency $\omega r_d / V_F$,
for $T / E_F = 0.01$, $f = 5.0$ (red squares); 
$T / E_F = 0.1$, $f = 5.0$ (green circles); 
and $T / E_F = 0.01$, $f = 3.0$ (black triangles, in this case 
$v_f$ is multiplied by factor $(5/3)^2$ to underline quadratic 
dependence on $f$). Here $R = 2$, $\theta = 0$.
}
\label{fig7}       
\end{figure}

The frequency dependence of $v_f/V_F$ is shown in Fig.~7.
The data presented there demonstrate that the frequency spectrum 
is independent of temperature (for $T \ll E_F$) and that quadratic 
dependence on $f$ is valid for a large frequency range.
The spectral dependence has a few characteristic features.
For $\omega r_d /V_F  \ll 1$ the ratchet velocity becomes independent 
of $\omega$, that is in agreement with the fact that 
$D_E \propto v_f$ is independent of $\omega$. This has been also 
seen in models analyzed in Refs.~\cite{second,cristadoro}.
For $\omega r_d /V_F  \gg 1$ the ratchet velocity drops 
with $\omega$. This is consistent with the dependence of 
$D_E$ on $\omega$ which in this limit can be estimated as 
$D_E \sim f^2 V_F^3 / \omega^2 l $. This comes from the fact 
that at high frequency the change of velocity after one collision
with semidisks is $\Delta v \sim f / \omega m$ so that 
the energy is changed by $\delta E \sim V_F f / \omega$ 
after a time  between two collisions $\delta t \sim l / V_F$ 
($l \sim r_d$ for $ R \sim r_d$ and $l \sim R^2/r_d$
for $R \gg r_d$). The frequency dependence shown
in Fig. 7 has also a resonance at $\omega r_d/V_F  \approx 3$. 
This is probably related to high frequency collisions 
in narrow bottle necks between two semidisks (see Fig. 1
where the narrow bottle neck is approximately by a factor of three 
smaller than disk radius).

\begin{figure}[t!] 
\vglue 0.5cm
\epsfig{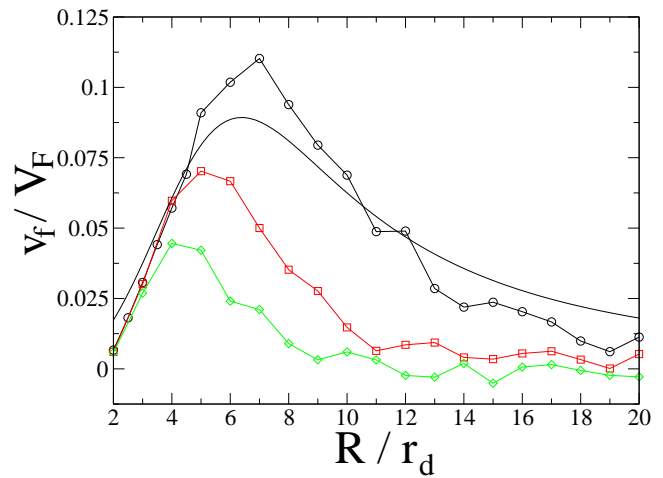}
\vglue -0.0cm 
\caption{Dependence of $v_f / V_F$ on the rescaled distance 
between semidisk centers $R / r_d$ for 
different frequency values $\omega = 1$, $1.5$, $2.0$
(curves from top to bottom). 
The smooth curve shows the fit of Eq.(2) for $\omega = 1$
with $A=0.017$, $B=0.012$. Here $f = 5.0$, $\omega = 1$, 
$\theta = 0$, $T / E_F = 0.1$. 
}
\label{fig8}       
\vglue +0.5cm
\end{figure}
 
The dependence of $v_f$ on the distance between disk centers $R$ 
is shown in Fig.~8. Initially $v_f$ starts to grow with $R$ then 
reaches a maximum value and drops at large $R$ values. 
The position of the maximum depends on the microwave frequency. 
With the increase of frequency the maximum moves to smaller values of $R$. 
Qualitatively this corresponds to the situation when the microwave frequency 
becomes comparable with the frequency of collisions of particles with 
semidisks. 
The dependence of data on parameters can be satisfactory described 
by a fit formula :
\begin{equation}
v_f / V_F  = A \frac{R^2 f^2}{E_F^2} \frac{1}{1 + B (\omega R^2 / r_d V_F)^2}
\end{equation} 
Here $A, B$ are dimensionless fitting parameters. 
The fit for three values of $\omega$ in Fig.~8 gives $A \approx 0.017$ 
and $B \approx 0.012$. For $\omega \rightarrow 0$ this expression 
is in a satisfactory agreement with the value $C \approx 0.13$ 
found in Fig.~6 at $R = 2$. The physical origin of this fit
is related to frequency dependence of the diffusion rate $D_E$ 
which interpolates between the low frequency regime 
($D_E$ independent of $\omega$) 
and the high frequency regime where $D_E$ drops quadratically with $\omega$
(see estimates given above). Eq.~(2) gives reasonable description of 
obtained numerical data in the regime where $l$ is not too large 
compared to $R$. This situation is most interesting for direct 
experimental studies where $R$ is not very large compared to $r_d$. 

Another important experimental parameter is the scattering time 
induced by impurities which are always present in real samples. 
The data presented in previous figures were obtained in the regime 
of very large $\tau_i$. The effect of finite values of $\tau_i$
on the ratchet velocity is described in Fig.~9 for various 
lattice constants $R$. 
The data show that at large $\tau_i$ values $v_f$ is independent 
of the impurity scattering time while at small $\tau_i$,
$v_f / V_F$ drops approximately linearly with $\tau_i V_F / r_d$. 
Indeed the asymmetry of semidisks is washed out by impurity 
scattering and the ratchet effect should disappear at small $\tau_i$. 
At the same time it is important to stress that the presence of 
impurities is not necessary for the onset of directed transport. 
In experimental conditions $\tau_i$ can depend on temperature 
because of electron-phonon scattering 
or electron-electron interactions which give a dependence of  
$\tau_i$ on $T$. This may lead to a significant
temperature dependence of the photogalvanic effect, in the
temperature range $T \sim 1^oK$.

\begin{figure}[t!]  
\vglue 0.5cm
\epsfig{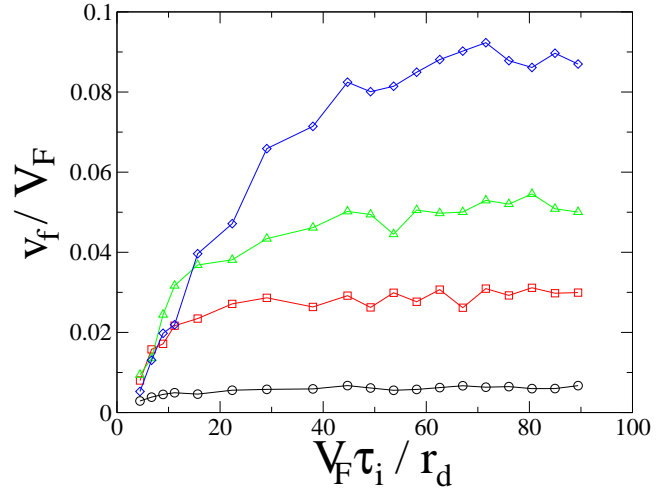}
\vglue -0.0cm 
\caption{(color online) Dependence of $v_f / V_F$ 
on the rescaled impurity scattering time $V_F \tau_i / r_d$ 
for $R / r_d = 6.0$, $4.0$, $3.0$, $2.0$ (from top 
to bottom respectively at $V_F \tau_i / r_d = 50$).
Here $f = 5.0$, $\omega = 1$, $\theta = 0$, 
$T / E_F = 0.1$.
}
\label{fig9}       
\vglue 0.5cm
\end{figure}

Experimentally it is also possible to study the dependence of the effect 
on magnetic field $B$ perpendicular to the 2DES plane.  
To investigate this dependence the method described above 
was adapted to the presence of a magnetic field, 
which was included in the analytical solution of motion equations 
between Metropolis thermalization steps. 
The magnetic field dependence is given in Fig.~10. 
The data clearly shows that the ratchet effect disappears
for sufficiently strong magnetic fields when the Larmor radius 
$r_l = V_F / B$ 
becomes smaller than the distance $R$ between semidisks (here 
electron charge and mass are set to $1$).
Indeed for $r_l \ll R$ the classical electron dynamics becomes 
integrable and the diffusion rate in energy $D_E$ goes to zero 
due to absence of chaos, thus leading to the disappearance of ratchet
($v_f \propto D_E$). In principle the magnetic field changes
the transport direction (angle $\psi$).
I do not discuss this dependence here
since the main point is that the ratchet effect disappears
at relatively low $B$ (see below).

\begin{figure}[t!]  
\vglue 0.5cm
\epsfig{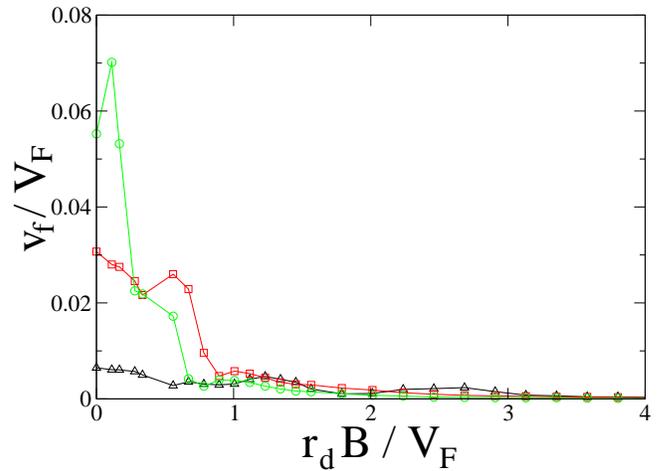}
\vglue -0.0cm 
\caption{(color online) Dependence of $v_f / V_F$ 
on the rescaled magnetic field $r_d B / V_F$
for $R / r_d = 4.0$, $3.0$, $2.0$ (curves from 
top to bottom at  $r_d B / V_F = 0$). 
Here $f = 5.0$, $\omega = 1$, $\theta = 0$, 
$T / E_F = 0.1$.
}
\label{fig10}       
\end{figure}

\section{Conclusions} 

The obtained results clearly shows the existence of zero 
mean force ratchet in asymmetric semiconductor structures
induced by microwave radiation. 
The obtained results give the following dependence for 
the strength of the stationary current induced by the ratchet effect 
in one row of semidisks (row width $\sqrt{3} R$): 
\begin{equation}
I = \sqrt{3} e n_e R v_f= 
A \sqrt{\frac{6}{\pi^2}} \frac{f^2}{n_e^{1/2}} \frac{e m R^3}{\hbar^3}
\end{equation}
where $E_F = \pi n_e \hbar^2 / m$ and $m=0.065m_e$.
This dependence holds in the low frequency regime which is 
usually satisfied at typical electron densities $n_e = 10^{12} cm^{-2}$ 
where the collision frequency is of the order of $200$GHz
for a $R \sim 1 \mu$m and under the assumption $R \sim r_d$
For $R = 1 \mu$m and $f / e = 1 $V/cm, the equation (3) 
gives the current $0.1$nA. In samples with high 
mobility the mean free path can have values as high as $5 \mu$m, 
and therefore the optimal regime for photogalvanic effect 
will be when $R$ is of the same order
and it is quite possible that in this situation the current per row 
can be as high as $10$nA. According to the results
of Fig.~10 for $R \sim 1\mu$m
the ratchet effect starts to disappear at magnetic field $B \sim 0.1 T$.

The asymmetric antidot lattice can be considered as a prototype 
for transport in asymmetric molecular structures. 
The latter have attracted recently a significant interest 
in view of possible biological applications
of ratchets \cite{ajdari2}.
Therefore experimental investigations on the ratchet effect 
discussed in this paper are highly desirable. 

I thank D. L. Shepelyansky for stimulating discussions 
and for his interest in this research. 
I am also grateful to M.~V.~Entin, and L.~I.~Magarill 
for access to their unpublished results on their kinetic
equation approach to the photogalvanic effect. 
This research is done in the frame of the ANR PNANO project MICONANO.


\begin{thebibliography}{99}
\bibitem{linke1} In Ref.~\cite{linke} the frequency of AC driving is 191 Hz; 
       this is much smaller than all relaxation rates and smaller than the 
       energy spacing between adjacent quantum levels inside one cell 
       $\Delta / \hbar \sim 1$GHz ; 
       therefore the regime discussed in this paper doesn't work for so low frequencies
       as discussed in \cite{cristadoro}.
\end{thebibliography}
\end{document}